\documentstyle[prd,preprint,aps]{revtex}

\newcommand{\fp}{f^{'}}

\begin{document}
\draft
\parskip 0.3cm
\begin{titlepage}

\begin{centering}
{\large \bf Limits on excited $\tau$ leptons masses from
   leptonic $\tau$ decays}\\

  \vspace{.9cm}
{\bf  Jorge Isidro Aranda}
\vspace{.05in}
\\ {\it Escuela de Ciencias F\'isico-Matem\'aticas,\\
Universidad Michoacana de San Nicolas de Hidalgo,\\
Morelia, Michoacan, Mexico.}\\

  \vspace{.4cm}
{\bf R. Martinez} \\
\vspace{.05in}
{\it  Departamento de F\'{\i}sica,
Universidad Nacional. \\
Bogota, Colombia.} \\
\vspace{.4cm}

{\bf O. A. Sampayo} \\
\vspace{.05in}
{\it  Departamento de F\'{\i}sica,
Universidad Nacional de Mar del Plata \\
Funes 3350, (7600) Mar del Plata, Argentina} \\ \vspace{.4cm}

\vspace{1.5cm}
{ \bf Abstract}
\\
\bigskip
\end{centering}
{\small
We study the effects induced by excited leptons on
the leptonic $\tau$ decay at one loop level. Using a general effective
lagrangian
approach to describe the couplings of the excited leptons, we compute
their contributions to the leptonic decays and use the current
experimental values of the branching ratios to put limits on the mass of
excited states and the subestructure scale.
  }
\pacs{PACS: 12.15.-y, 12.60.Rc, 13.35.Dx}

\vspace{0.2in}
\vfill

\end{titlepage}
\vfill\eject


The values for the leptonic $\tau$ decays \cite{pdgp}
have confirmed
the vality of the standard model (SM) \cite{weinberg} as the theory
of electroweak
interaccions at the current scales of energies. Morever, the results have
reached such precision that they have opened the possibility to constrain
significantly some physics beyond the standard model, for instance,
compositeness \cite{harari,loreyo}.

The family structure of the known fermions, among other regularities, has
been considered as an indication to expect that the SM fermions and
perhaps massive gauge bosons possess some kind of substructure. The idea of
composite models assumes the existence of an underliying structure,
characterized by scale $\Lambda$, with the fermions sharing some of the
constituents. As a consequence, excited states of each
known lepton should show up at some energy scale, and the SM should be seen
as the low-energy limit of a more fundamental theory.

Precise measurements of anomalous magnetic moment of muon and electron 
indicate that
first and second family of leptons are elementary particles with high grade 
of precision.
In this conditions, for simplicity, we take an conservative point of view
and we assume that only $\tau$, $\nu_\tau$ can be composite. They are
largely
most massive that the others leptons and their properties are less known.
Then, in this work, we consider $\tau$ and $\nu_\tau$ leptons as composite 
and
we keep the other leptons as elementary. In this conditions we only 
consider
excited states of leptons $\tau$ and $\nu_\tau$. It is our fundamental
hypotesis that can be understood considering either that the first and 
second
family
are elemtary or their asociate subestructure scale are much bigger than the
$\tau$ compositeness scale ($\Lambda_e,\Lambda_{\mu} >> \Lambda_{\tau}$).

We still do not have a satisfactory model, able to
reproduce the whole particle spectrum. Due the lack of a predictive theory
we should rely on a model-independent approach to explore the possible
effects of compositeness, employing effective Lagrangian techniques
to describe the couplings of these states.

Several experimental collaborations have been searching for excited states
\cite{experimen},
in particular on $\tau^*$ and $\nu^*_{\tau}$.
Their analyses are based on an effective SU(2)$\otimes$U(1) invariant
Lagrangian, proposed some years ago by Hagiwara et. al.\cite{hagiwara}.
Also a
  series of phenomelogical studies of excited fermions have been carried
out in several experiments. Moreover theorical bounds have been derived
from the contribution to the anomalous
magnetic moment of leptons and the Z scale observables at LEP.

On the other hand, an important source of indirect information about new
particles and interactions is the precise mesurement of the leptonic
branching ratios (BR) of lepton $\tau$ \cite{loreyo}.
Virtual effects of these new states
can modify the SM predictions for the BR, and the comparison with the
experimental data can impose bounds on their masses and couplings.

In this paper we use a general effective Lagrangian approach 
\cite{hagiwara}
to investigate the effects induced by excited tau and tau neutrinos
in the leptonic branching ratios at one loop level. We show our
results as an allowed region in the ($m^*,f/\Lambda$) plane. We find
bounds for
the subestructure scale as a function of excited mass and compare them with
bounds obtained for different experiments, in particular OPAL
  \cite{experimen}, bounds coming from the anomalous weak-magnetic
moment of the tau lepton\cite{gabriel,mconcha} and precision mesurement on 
the
Z peak \cite{mconcha1}.

The SM prediction for the leptonic decay width, including electroweak
radiative corrections, is

\begin{equation}
\Gamma_{SM}(\tau\rightarrow l \nu_l \nu_{\tau})=\frac{G_F^2 m_{\tau}^5}{192 
\pi^3}
f(m_l^2/m_{\tau}^2)r
\end{equation}
where
$$f(x)=1-8x+8x^3-x^4-12x^2\ln(x)$$
and the factor r takes into account the radiative corrections that are not
absorbed in the Fermi constant $G_F$, and is estimated to be 0.9960.

The leptonic branching can be calculated in terms of the observed $\tau$
lifetimes, $\tau=(290.0\pm1.2)\times10^{-15}$s, and the measured value for
the mass of the $\tau$ lepton, $m_{\tau}=1777.0\pm0.28$ MeV. With these, it 
is possible to stimate the theoretical values  for the branching
ratio of the electronic and the muonic modes: $B_e^{th}=0.1777\pm0.0007$ 
and $B_{\mu}^{th}=0.1728\pm0.0008$, respectively which are now in good 
agreement whith the experimental results:
$B_e^{exp}=0.1781\pm0.0007$ and $B_{\mu}^{exp}=0.1737\pm0.0009$, once the
theoretical uncertainties are properly taken into account \cite{pdgp}.
We have also considered the quantity $R_\tau$, which is defined as

\begin{equation}
R_\tau=\frac{\Gamma_\tau-\Gamma_e-\Gamma_\mu}{\Gamma_e} \; .
\end{equation}
Using the measured values of the leptonic branching ratios, one find the
value

\begin{equation}
R^{exp}_\tau=\frac{1-B_e-B_\mu}{B_e}=3.64\pm0.019
\end{equation}

In order to study limits on the scale of compositeness, we shall consider
the contribution to the decay width, due to indirect effects induced by
excited $\tau^*$ and $\nu^*_{\tau}$ at one loop level. For hypothesis
the other leptons
are considered either elementary or with their excited states decoupled
due to  much bigger composinteness scale. We consider excited
fermionic states with spin and isospin $\frac12$, and we assume that
the excited fermionic acquire masses before the $SU(2)\times U(1)$
breaking, so that both left-handed and right-handed states belong
to weak isodoublets (vector-like model).
The effective dimension five Lagrangian
that describes the coupling of excited-usual fermions, which is
SU(2)$\times$U(1), can be written as \cite{mconcha}
\begin{eqnarray}
{ \cal L }_{eff}=&-&\sum_{V=\gamma,Z,W} T_{VLl} \overline L \sigma^{\mu\nu}
P_L l\partial_{\mu}V_{\nu}   \nonumber \\
&-&i \sum_{V=\gamma,Z}
Q_{VLl} \overline L \sigma^{\mu\nu} P_L lW_\mu V_\nu + h.c.
\end{eqnarray}
where $L=\nu_{\tau}^*,\tau^*$ represent the excited states,
and $l=\nu_\tau,\tau$, the usual
light fermions of third generation. A pure left-handed structure is
assumed for these couplings.
The coupling constants $T_{VLl}$ are given by
\begin{eqnarray}
T_{\gamma \tau^* \tau}&=&-\frac{e}{2\Lambda}(f+\fp) \nonumber\\
T_{\gamma \nu^*\nu}&=&\frac{e}{2\Lambda}(\fp-f)  \nonumber\\
T_{Z\tau^*\tau}&=&-\frac{e}{2\Lambda}(\fp \cot\theta_W-f \tan\theta_W) \\
T_{Z\nu^*\nu}&=&\frac{e}{2\Lambda}(\fp \cot\theta_W+f \tan\theta_W)
\nonumber \\
T_{W\tau^*\nu}&=&T_{W\nu^*\tau}=\frac{e}{\sqrt{2}\sin\theta_W\Lambda}\fp
\nonumber,
\end{eqnarray}
where $\Lambda$ is the compositeness scale, $\fp$ and $f$ are weight 
factors
associated to the $SU(2)$ and $U(1)$ coupling constants and $\theta_W$ is 
the
weak mixing angle.
The quartic interaction couplings
$Q_{VLl}$ are given by,
\begin{eqnarray}
Q_{\gamma \tau^* \nu}=-Q_{\gamma \nu^* \tau}=
-\frac{e^2\sqrt2}{2\sin\theta_W\Lambda} \fp  ,\\
Q_{Z\tau^*\nu}=-Q_{Z\nu^*\tau}=
-\frac{e^2\sqrt2cos\theta_W}{2\sin^2\theta_W\Lambda} \fp .\nonumber
\end{eqnarray}

The coupling of gauge bosons to excited leptons in a vector-like model are
given by the following renormalizable
lagrangian (dimension four),
\begin{eqnarray}
{\cal L}_{ren}=-\sum_{\gamma,Z,W} A_{VLL} \overline L\gamma^{\mu} V_\mu L
\end{eqnarray}
which is $SU(2)\times U(1)$ invariant. The coupling constants are given by,
\begin{eqnarray}
A_{\gamma \tau^*\tau^*}&=&-e  \;\; , \nonumber\\
A_{\gamma \nu^*\nu^*}&=&0 \;\; ,\nonumber \\
A_{Z\tau^*\tau^*}&=&\frac{(2\sin^2\theta_W-1)e}{2\sin\theta_W\cos\theta_W}
\;\; ,\\
A_{Z\nu^*\nu^*}&=&\frac{e}{2\sin\theta_W\cos\theta_W}\;\; , \nonumber \\
A_{W\tau^*\nu^*}&=&\frac{e}{\sqrt2\sin\theta_W} \;\; . \nonumber
\end{eqnarray}

The contributions of the excited leptons to amplitude for the
leptonic tau decay at one-loop level are represented in figure 1. We
consider the tree level amplitude plus the legs, box and vertex corrections
which are representd in figure 2 and 3, respectively.
The dominant contributions from this radiative corrections are given by the
interference between the SM term and the new contributions from excited
$\tau^*$ and $\nu^*_\tau$. We find that the decay width can be written as
\begin{equation}
\Gamma=\Gamma_{SM} ( 1 + \delta\Gamma^{(RC)})
\end{equation}
where the SM part is given by Eq(1), and the expression for the new part,
taking $m_\tau^*=m_{\nu_\tau}^*=m^*$, is
\begin{equation}
\delta\Gamma^{(RC)}=\frac{\alpha}{f(m_l^2/m_{\tau}^2) r}
\left(\frac{m^*f}{\Lambda}\right)^2 ({\cal L}+{\cal V}+{\cal B}).
\end{equation}
where $\alpha$ is the fine structure constant. The functions
${\cal L}$, ${\cal V}$ and ${\cal B}$ correspond to the
interference between the SM term and the Leg, Vertex and Box radiative
corrections, respectively.

The functions
${\cal L}$, ${\cal V}$ and ${\cal B}$ are given by
\begin{eqnarray}
{\cal L} &=& -\frac{3}{4\pi}\left[-\frac{c_w^6\xi_z^6
\ln(c_w^2\xi_z)}{s_w(1-c_w^2 \xi_z)^2}
-\frac{(c_w^4+s_w^4)\xi_z^3 \log(\xi_z)}
{2s_w^2c_w^2(1-\xi_z)^2} + \frac{11}{3}   \right.
\nonumber \\
&+&\left(2+\frac{2+c_w^2\xi_z}{s_w^2}
+ \frac{(2+\xi_z)(c_w^4+s_w^4)}{2c_w^2s_w^2}\right)\ln\!\xi_{\Lambda}
\\ \nonumber
&+&\left.\frac{(c_w^4+s_w^4)(22-11\xi_z-17\xi_z^2)}{12 c_w^2s_w^2(1-\xi_z)}
+\frac{(22-11c_w^2\xi_z-17c_w^4\xi_z^2)}{6s_w^2(1-c_w^2\xi_z)}
\right] ,
\nonumber \\ \nonumber
\end{eqnarray}
\begin{eqnarray}
{\cal V} &=& \frac{1}{16\pi s_w^2 \xi_z(1-\xi_z)^2(1-c_w^2 \xi_z)^2}\left[
-24 c_w^4 \xi_z^3(1-\xi_z)^2 \ln(c_w^2\xi_z)\right.
\nonumber \\ \nonumber \\
&+&6(1-c_w^2\xi_z)\xi_z^3\left(4(1-\xi_z)+
s_w^2\xi_z\left(\frac{c_w^2}{s_w^2}-\frac{s_w^2}{c_w^2}+\frac{4}{c_w}\right)
\right)\ln(\xi_z)
\nonumber \\ \nonumber \\
&+&6(1-c_w^2\xi_z)s_w^2\xi_z(1-\xi_z)^2\left((2+\xi_z)
\left(4-\frac{2}{c_w}-\frac{c_w^2}{s_w^2}+\frac{s_w^2}{c_w^2}\right)\right.
\\ \nonumber \\
&+&\left.8\left(\frac{1}{s_w}-1\right)\right)\ln(\xi_{\Lambda})+
(1-c_w^2\xi_z)(1-\xi_z)s^2\xi_z\left(20\xi_z(1-\xi_z)\right.
\nonumber \\ \nonumber \\
&+& \left.\frac{12}{c_w} \xi_z(1+\xi_z)+
\left.\left(\frac{c_w^2}{s_w^2}-\frac{s_w^2}{c_w^2}\right)(2+5\xi_z-\xi_z^2)\right)
\right] ,
\nonumber \\ \nonumber
\end{eqnarray}
\begin{eqnarray}
{\cal B} &=& \frac{9(1-3s_w^2)\xi_z}{8\pi s_w^2}
\left[\frac{1}{36s_w^2(1-\xi_z)(1-c_w^2\xi_z)}
\left(11(1-\xi_z)(1-c_w^2(1+\xi_z)
\right. \right.
\nonumber \\ \nonumber \\
&+& c_w^4\xi_z)+6\xi_z(1-c_w^2\xi_z-c_w^4(1-\xi_z))\ln(\xi_z)
\\ \nonumber \\
&-&\left. \left. 6c_w^4\ln(c_w^2)\xi_z(1-\xi_z)\right)
\frac{\ln(\xi_\Lambda)}{6}-\frac{43-26c_w^2}{72(1-3s_w^2)}\right]
\nonumber
\end{eqnarray}
where $\xi_z=(m_Z/m^*)^2$, $\xi_w=(m_W/m^*)^2$,
$\xi_{\Lambda}=(m_{\Lambda}/m^*)^2$, $s_w=\sin\theta_w$ and 
$c_w=\cos\theta_w$.
In the following analysis we take $m_\tau^*=m_{\nu_\tau}^*=m^*$
and $f=\fp$.

The loops contributions of the excited leptons were evaluated in
$D=4-2 \epsilon$
dimensions using the dimension regularization method, which is a 
gauge-invariant
regularization procedure, where the pole at $D=4$ is identified with
$\ln \Lambda^2$.

We should notice that since we are including non-renormalizable
operators the results of the loops are, in principle,
quadratically divergent with the scale $\Lambda$. However, since we are
restricting ourselves to $SU(2) \times U(1)$ gauge invariant operators,
the final results for the physical observables are, at most, 
logarithmically
divergent. In other words, all quadratic (or higher) dependence on
$\Lambda$ is simply cancelled by counter terms coming from the high-energy
theory. At one loop at best only a logarithmic dependence on the scale of
new physics can be extracted purely from the low-energy effective 
lagrangian.

We evaluate the renormalization constants by imposing the on-shell
renormalization conditions on the renormalized transition amplitudes.
In this scheme we compute the diagrams of the externals legs to obtain the
renormalization of the lepton wave functions taking the mass of the
particles as the experimental value.

To obtain bounds for the excited mass and the subestructure scale we 
compare
our theoretical results: $B_e^{th}=\Gamma_e^{th}/\Gamma_{\tau}$ (where
$\Gamma_e^{th}$ is given by eq.(9)) with the experimental value of
$B_e^{ex}$, $B_{\mu}^{ex}$ and $R_{\tau}$. We find that the most 
restrictive
bound come from $B_e^{ex}$ and then we use it to put limits on $\Lambda$
and $m^*$.
We consider $B_e^{th}$ as a function of $m^*$ and $\Lambda$
and then the limits are obtained by comparing $B_e^{th}(m^*,\Lambda)$ with 
the
experimental value of $B_e^{ex}$.

In Fig. 4 we show our bounds showing the allowed regions
for the excited lepton masses and the ratio $f/\Lambda$ at $95 \%$ C.L.
The curve that limit the region is obtained intersecting the function
$B_e^{th}(m^*,\Lambda)$ with the experimental value
$B_e^{ex}\pm\Delta B_e$.
It is understood that the non-allowed region sets co-related bounds
for the excited lepton mass and $\Lambda$.
To compare our results with other bounds we include
in this figure the results from OPAL collaboration and bounds coming from 
the
weak-magnetic moment of the $\tau$ lepton on the Z
peak \cite{experimen,gabriel,mconcha}.
Moreover in figure 5 we include a comparation between ours results and
bounds coming from precision mesurements on the Z peak \cite{mconcha1}.
It is important observer that the bounds from the leptonic tau decay
is safe in the $\Lambda > m^*$ region where the decoupling of new
physics work.

Finally we study the decoupling properties of the new contributions which
cancel out in the limit of large subestructure scale and fixed excited 
mass.
The results are shown in figure 6. The curves represent the variation of 
the
new contributions ($\delta \Gamma^{RC}$)with $\Lambda$ for
differents values of $m^*$.

Summarizing, we have considered the possibility that the lepton tau and
their neutrino have some kind of substructure. We have modeled the
interactions involving their excited states through a renormalizable
lagrangian (vector-like model) and an effective dimensi\'on-5 operator
that couples ordinary particles with excited particles and gauge bosons
and we have considered that the other leptons are either elementary or
their compositeness scale are much bigger than the tauonic one. By
computing the contributions of these interactions to the radiative 
corrections
of leptonic tau decay and by comparing them with the well mesured branching
ratios, we have obtained bounds on the excited state masses and the
compositeness scale $\Lambda$. This bounds are most restrictives that 
others
obtained from direct productions \cite{experimen}, from radiative 
corrections to
weak-magnetic moment of the $\tau$ lepton on the Z
peak \cite{gabriel,mconcha} and from precision measurements on the Z peak
\cite{mconcha1}.

{\bf Acknowledgements}

We thank to COLCIENCIAS (Colombia), CONACyT (Mexico) and CONICET 
(Argentina)
for their financial supports.

\pagebreak

\noindent{\large \bf Figure Captions}\\

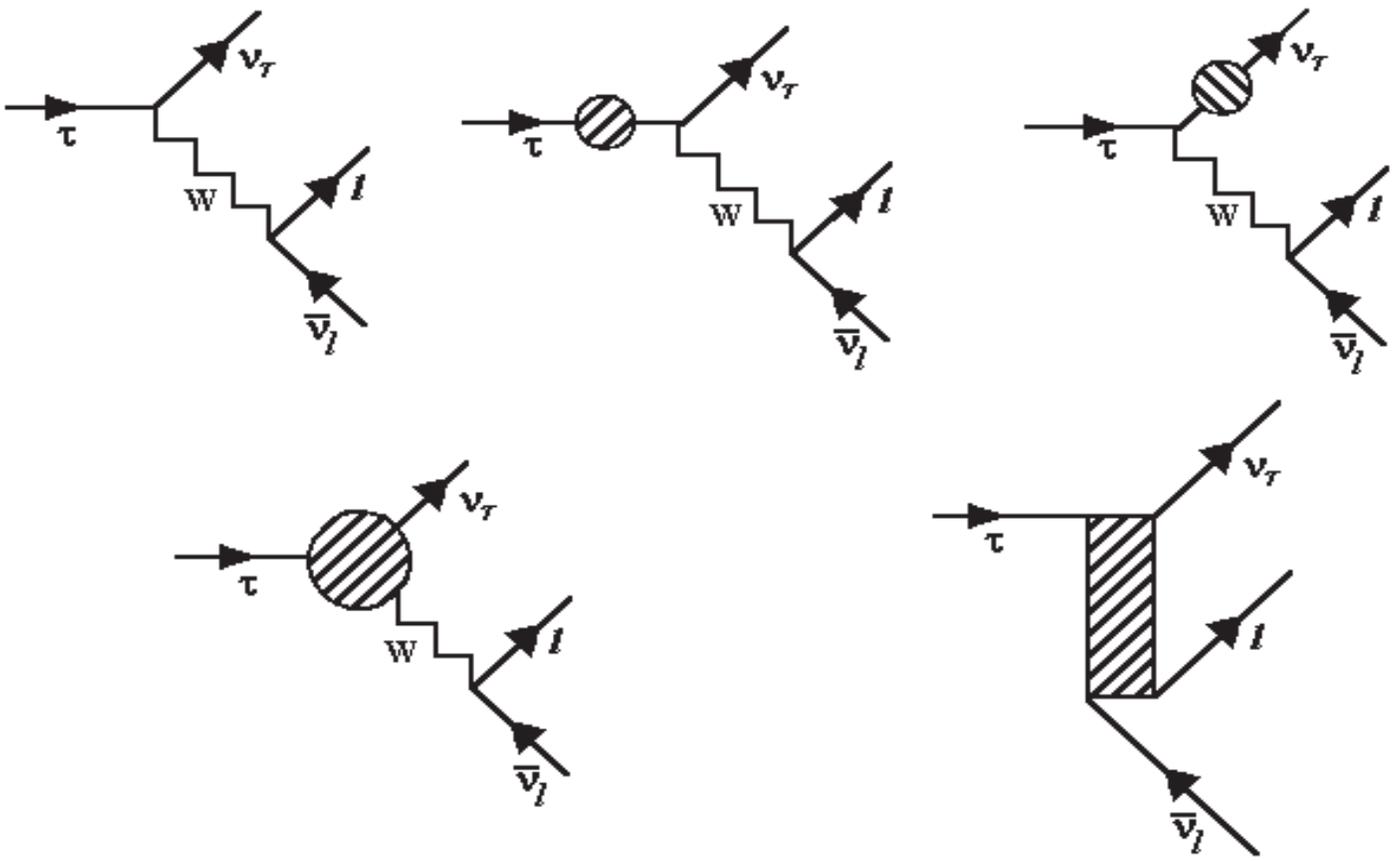
\noindent{\bf Figure 1:} Diagramatic representation for the contribution of 
the
leptonic tau decay amplitude. Box and dashases blobs represent the
contributions of excited tau and tau neutrino.

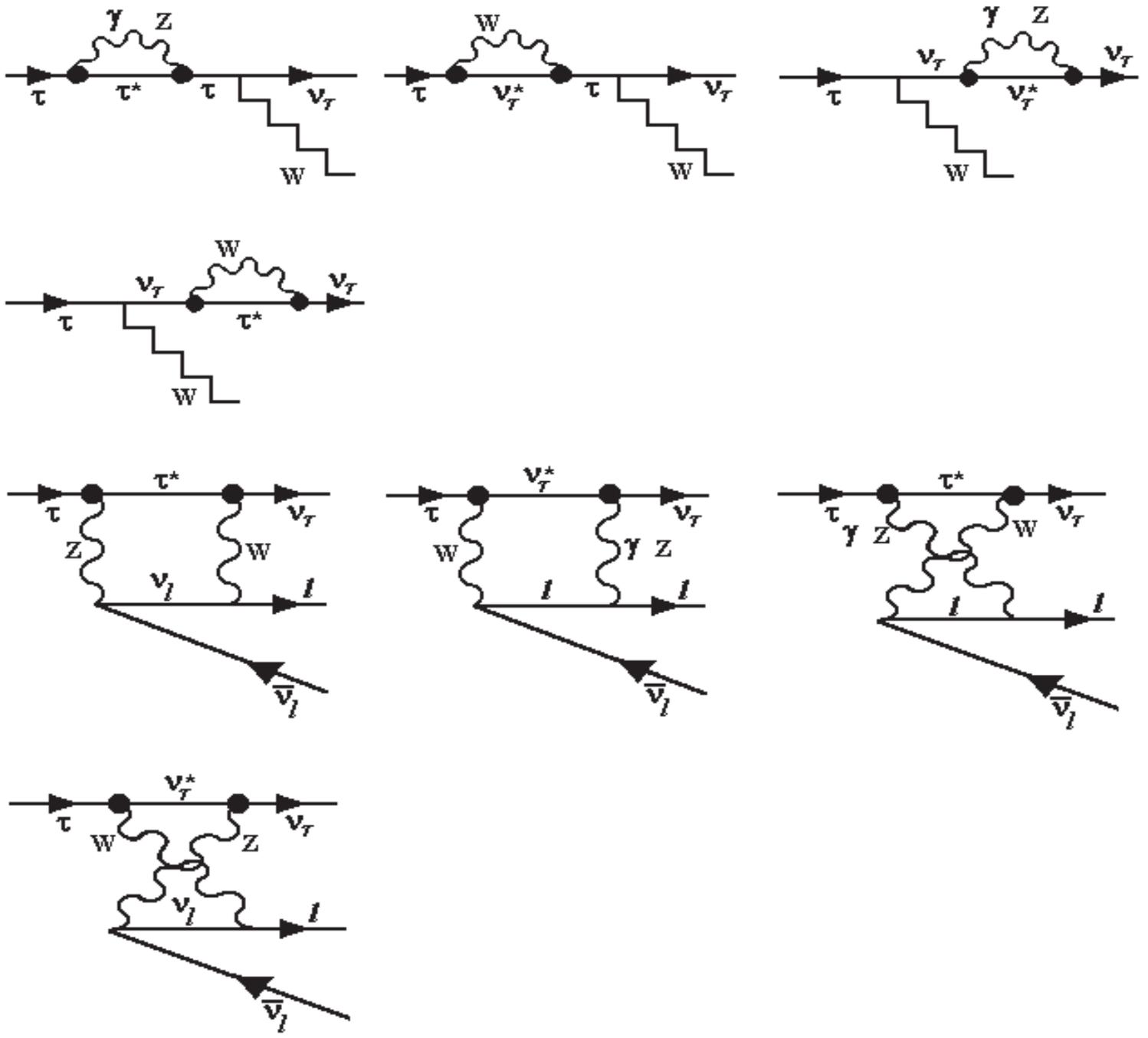
\noindent{\bf Figure 2:} Self-energy and box corrections contributing to
leptonic tau decay.

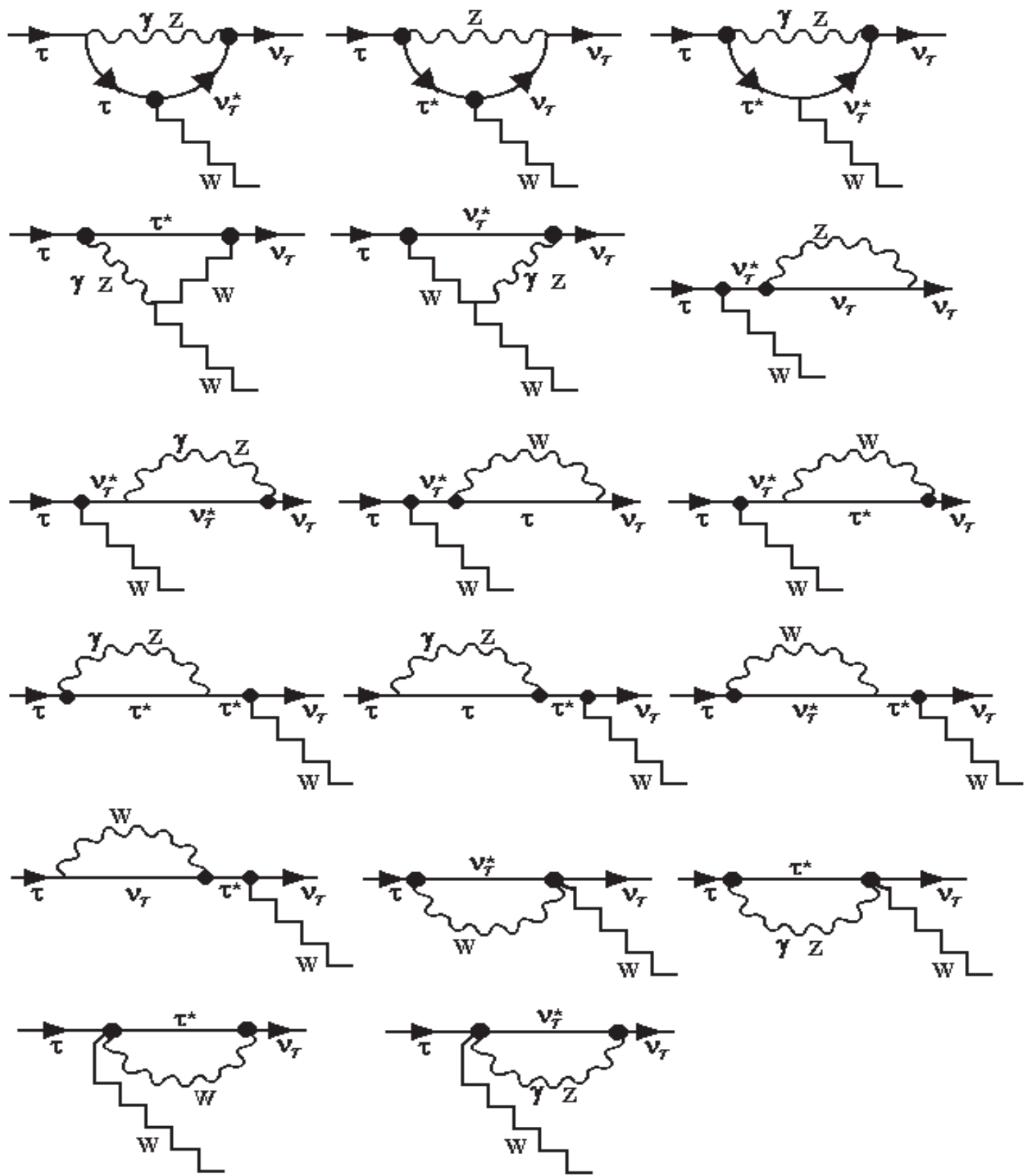
\noindent{\bf Figure 3:} Vertex corrections contributing to
leptonic tau decay.

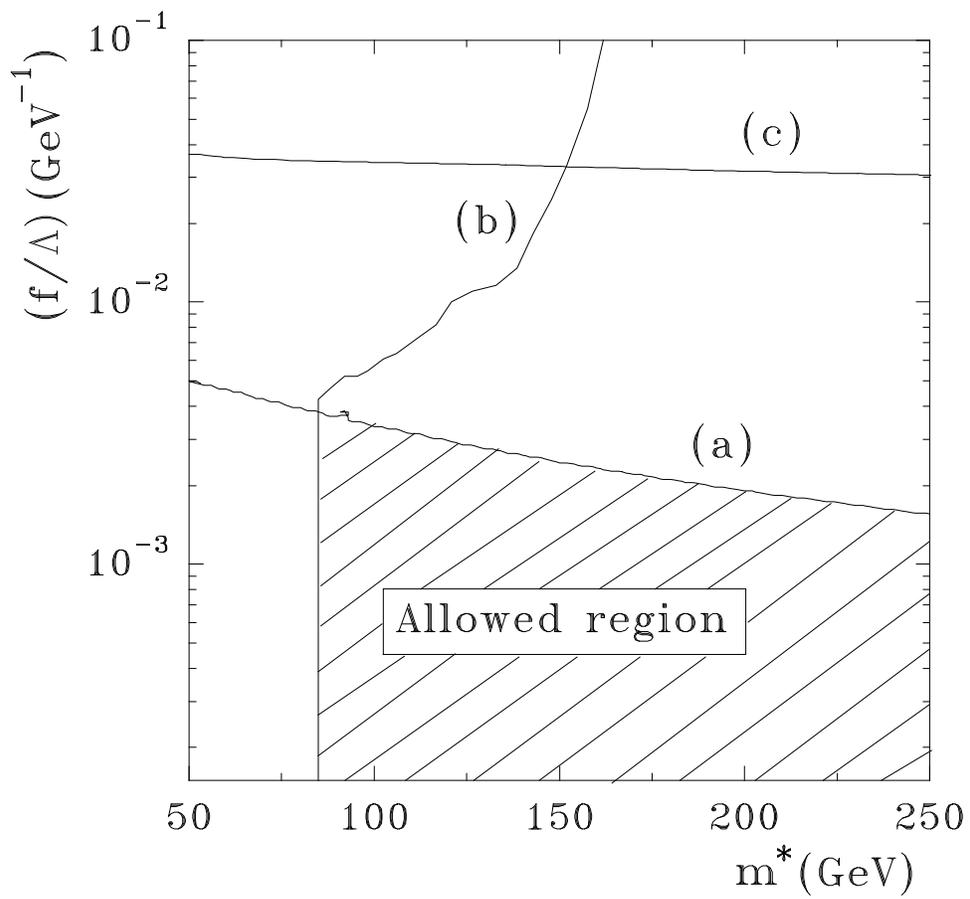
\noindent{\bf Figure 4:} Dashed zone repesent the allowed region at 95\% 
C.L.
The curves represent bounds coming from (a) leptonic tau decay,
(b) single production at OPAL and (c) weak-magnetic moment
of tau lepton.

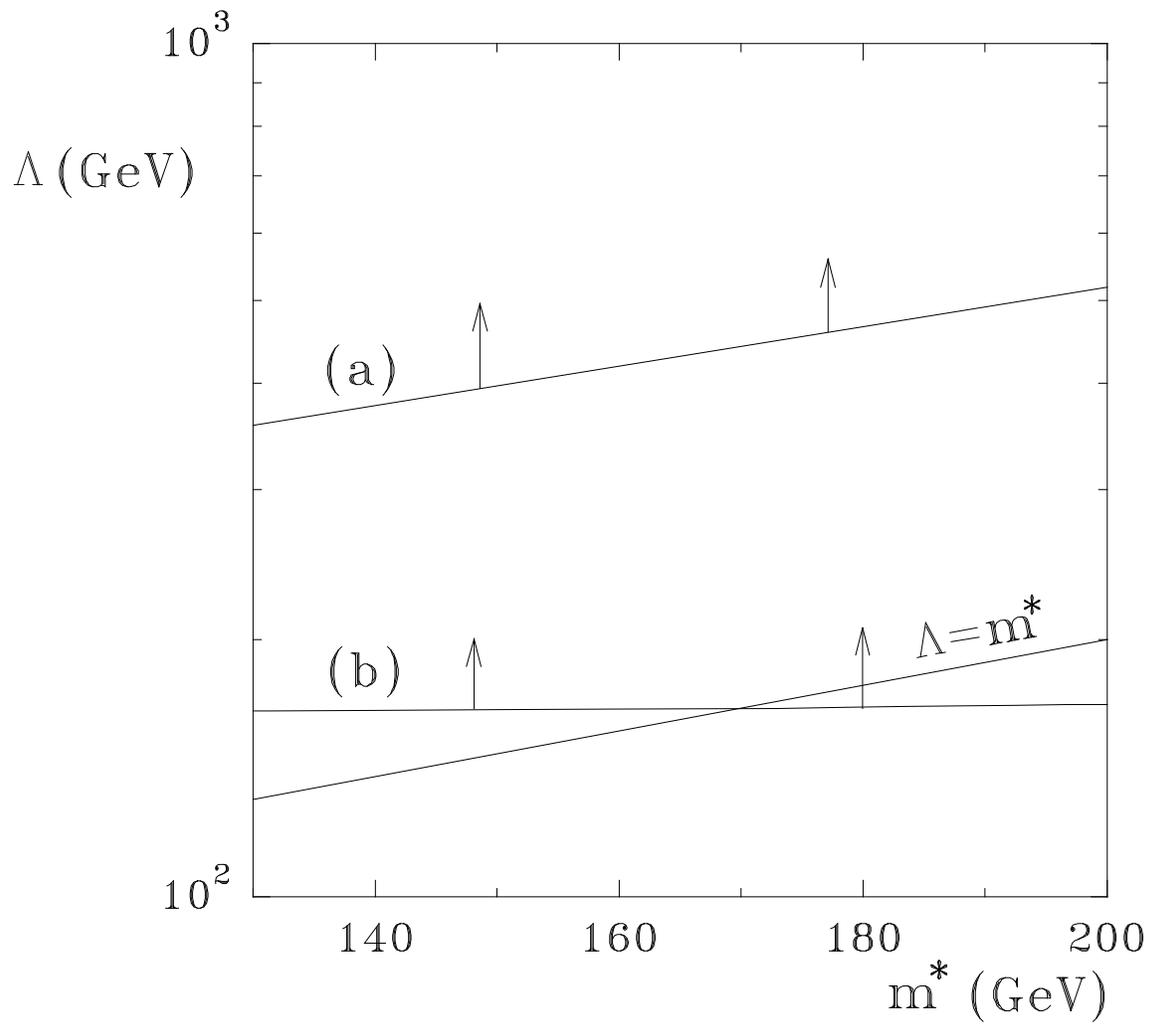
\noindent{\bf Figure 5:} Excluded regions in the $\Lambda$ versus $m^*$
plane (below of the curves), at 95 \% C.L., from (a) leptonic tau decay
and (b) precision mesurements on the Z peak.

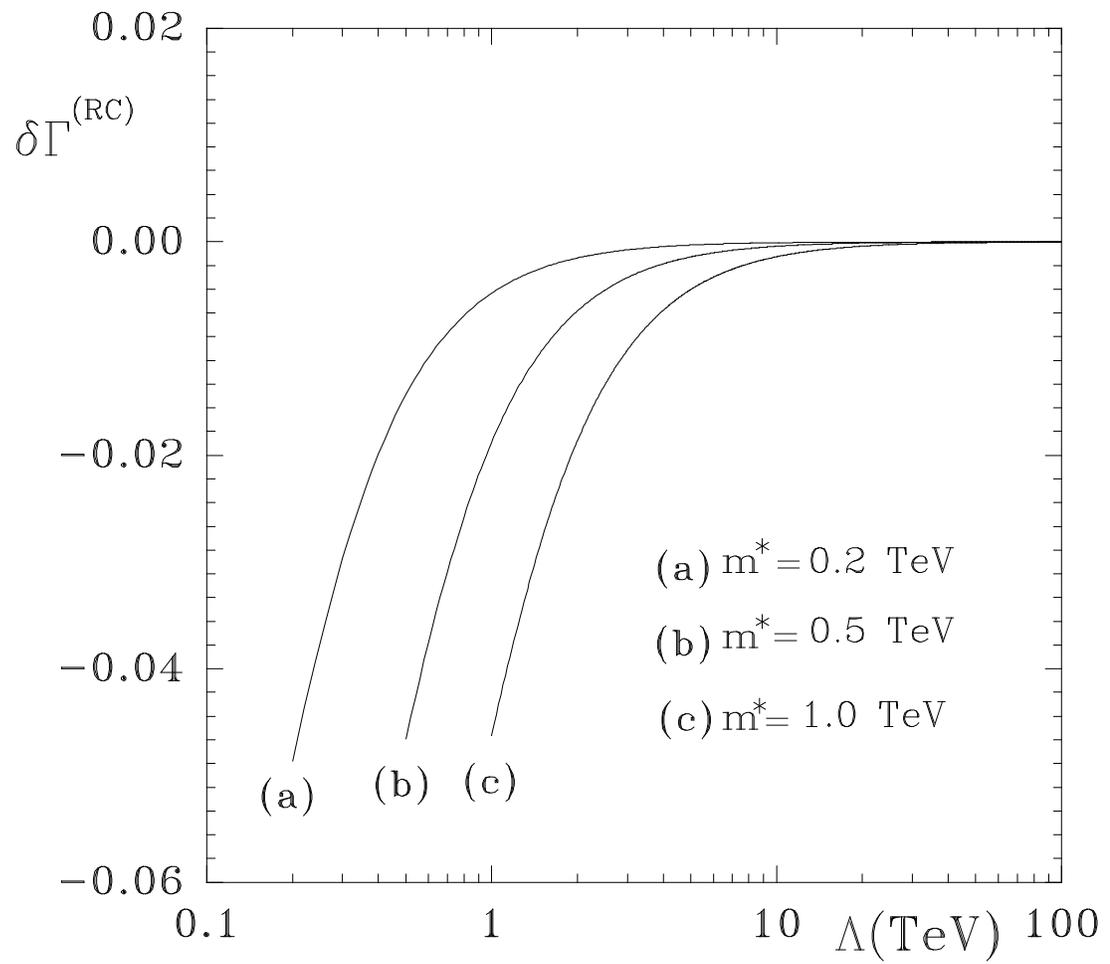
\noindent{\bf Figure 6:} Decoupling properties of the new contributions
as a function of $\Lambda$ for differents values of $m^*$.

\end{document}